\documentclass{pj}
\usepackage{graphicx}
\usepackage{amsmath,amssymb}

\newcommand{\etal}{\textit{et al}.}
\newcommand{\ie}{\textit{i}.\textit{e}.}

\begin{document}
\setcounter{page}{1}
\pjheader{July 2, 2020}

\title[Rigorous Quantum Formulation of Parity-Time Symmetric Coupled   Resonators]
{Rigorous Quantum Formulation of Parity-Time Symmetric Coupled   Resonators}  \footnote{\hskip-0.12in*\, Corresponding
author:~Shaolin~Liao~ (sliao5@iit.edu).} 
\footnote{\hskip-0.12in\textsuperscript{1} S. Liao is with Department of Electrical and Computer Engineering, Illinois Institute of Technology, Chicago, IL 60616 USA. \textsuperscript{2} L. Ou (oulu9676@gmail.com) is with College of Computer Science and Electronic Engineering, Hunan University, Changsha, Hunan, China  410082.}

\author{Shaolin~Liao\textsuperscript{*, 1} and Lu~Ou\textsuperscript{2}}

\runningauthor{Liao and Ou}

\tocauthor{FistName1~LastName1 and FistName1~LastName1}

\begin{abstract}
Rigorous quantum formulation of the Parity-Time (PT) symmetry phenomenon in the RF/microwave regime for a coupled coil resonators with lump elements has been presented. The coil resonator is described by the lump-element model that consists of an inductor (L), a resistor (R) and a capacitor (C). Rigorous quantum Hamiltonian for the coupled LRC coil resonators system has been derived through twice basis transforms of the original basis. The first basis transform rotates the original basis such that off-diagonal terms of the governing matrix of the equation system of the coupled coil resonators reduces to constants. Then a second basis transform obtains the quantum Hamiltonian, including the diagonal effective complex frequencies and the off-diagonal coupling terms, together with the transformed basis. With the obtain quantum Hamiltonian, the eigenvalues and eigenvectors of the coupled coil resonators can be obtained as usual as the quantum Hamiltonian. Finally, numerical simulation verifies the correctness of the theory. The quantum formulation of the coupled coil resonators can provide better guideline to design a better PT-symmetric system.
\end{abstract}


\setlength {\abovedisplayskip} {6pt plus 3.0pt minus 4.0pt}
\setlength {\belowdisplayskip} {6pt plus 3.0pt minus 4.0pt}

\

\section{Introduction}
\label{sec:intro}
Parity-Time (PT) symmetry has shown great potential as ultra-sensitive sensors \cite{hajizadegan_high-sensitivity_2019, sakhdari_ultrasensitive_2018}, Wireless Power Transfer (WPT \cite{paul_fast_2018, assawaworrarit_robust_2017}), gain/loss controlled lasers \cite{gao_parity-time_2017} and absorber \cite{wang_coherent_2019}, in both the photonics \cite{phang_localized_2016, cui_exceptional_2019}  and microwave \cite{liu_observation_2018-1} regimes. PT symmetry is a physics phenomenon that originates from the quantum community: it was first proposed in quantum mechanics by Bender and Boettcher in 1998 \cite{bender_real_1998}. Counterintuitively, it is argued that real eigenfequencies exist even for a non-Hermitian Hamiltonian when a quantum system is invariant under operations of spatial reflection $\mathcal{P}(x)$ and time reversal $\mathcal{T}(t)$ operations. The existence of PT-symmetric quantum system has not been experimentally demonstrated due to the difficulty of generating the non-Hermitian quantum system. However, gain and loss can be readily introduced in photonics \cite{liao_optimal_2020, liao_optimal_2019_2, liao_optimal_2019, peng_zim_2019, liao_high-q_2020, liao_miniature_2018, peng_-chip_2019,  zeng_integrated_2019}. So PT symmetry can be introduced by designing the proper gain-loss profile so that the PT symmetry phenomenon happens. Following the pioneering theoretical work by El-Ganainy \etal   \cite{el-ganainy_theory_2007}, the feasibility of translating this quantum-inspired symmetry to the optics regime has been demonstrated in a various contributions and, specifically, in coupled optical structures \cite{makris_beam_2008, musslimani_optical_2008, ruschhaupt_physical_2005, klaiman_visualization_2008, longhi_time_2017, rodriguez-lara_symmetry_2018}. Later, it has been also demonstrated in the electromagnetic and acoustic systems \cite{hodaei_parity_timesymmetric_2014, savoia_symmetry_induced_2015}, whose governing Helmholtz equation \cite{liao_image_2006, liao_near-field_2006, shaolin_liao_new_2005, liao_fast_2006, liao_cylindrical_2006, liao_beam-shaping_2007, liao_fast_2007, liao_validity_2007, liao_high-efficiency_2008, liao_four-frequency_2009, vernon_high-power_2015, liao_multi-frequency_2008, liao_fast_2007-1, liao_sub-thz_2007, liao_miter_2009, liao_fast_2009, liao_efficient_2011, liao_spectral-domain_2019} is similar to the Schr\"{o}dinger equation in the quantum physics. These PT-symmetric wave systems are usually realized by introducing spatial distribution of balanced gain-loss profiles.

In particular, RF and microwave frequencies, a PT-symmetric system can be readily realized with transmission-line networks or lumped-element circuits \cite{schindler_experimental_2011, schindler_symmetric_2012, radi_parity_time_symmetric_2016, chen_generalized_2018}.  However, most of the works above take the quantum Hamiltonian as granted, assuming that the diagonal natural frequencies are the resonant frequencies of the coupled coil resonators and the off-diagonal coupling terms are known. Also, it generally assumes that the basis of the quantum Hamiltonian consists of stored energies in the coupled coil resonators, but no clear formula exists. In this paper, rigorous quantum formulation has been derived and all of these will be clear, providing guideline to design a better coupled coil resonators and its system \cite{hajizadegan_high-sensitivity_2019, zhu_compact_2018, padmaraj_compact_2019}.

\section{The Coupled LRC Coil Resonators}
\label{sec:formulation}
A  resonant coil can be modeled as a LRC tank that consists of 3 components in parallel: a inductor with inductance L, a capacitor with capacitance C and a resistor R. 

\subsection{The Governing Equations System}
When a pair of  LRC resonant tanks are brought close to each other, they are coupled together through magnetic flux of the two inductive coils, which can be characterized by the mutual inductance $M$. The coupled resonant coils can be analyzed by the physical quantities of currents $i_1/i_{2}$ and voltages $v_{1}/v_{2}$ through the indicators of the two coupled coil resonators, 
\begin{align}\label{eqn:IV}
v_{1} = L_1 \frac{d i_1}{dt}  + M \frac{d i_{2}}{dt}, \ \
v_{2} = L_{2} \frac{d i_{2}}{dt}  + M \frac{d i_{1}}{dt}. 
\end{align}

The Kirchhoff Current Law (KCL) connects the currents through the inductor with inductance $L$, the capacitor with capacitance $C$ and the resistor with conductance $G$ as follows,
\begin{align}\label{eqn:KCL}
i_{1}  + C_1 \frac{dv_1}{dt} + G_1 v_1 = 0, \ \  i_{2}  + C_2 \frac{dv_{2}}{dt} + G_{2}   v_{2}  = 0.   
\end{align} 


Taking Fourier transforms on Eq. (\ref{eqn:IV}) and Eq. (\ref{eqn:KCL}), after some mathematics, the following are obtained,
\begin{align}\label{eqn:IV_v_Laplace}
\left(\frac{-\omega^2}{\omega_1^2} + \frac{j\omega}{\alpha_1 } + 1\right) v_{1}(\omega) + \left(\frac{-\omega^2}{\kappa_{2}^2} + \frac{j\omega}{\gamma_{2}} \right) v_{2}(\omega) =0 \\ 
\left(\frac{-\omega^2}{\omega_{2}^2} + \frac{j\omega}{\alpha_{2} } + 1\right) v_{2}(\omega) + \left(\frac{-\omega^2}{\kappa_{1}^2} + \frac{j\omega}{\gamma_{1}} \right) v_{1}(\omega) =0, \nonumber
\end{align}
with
\begin{gather}
\omega_1 = \frac{1}{\sqrt{L_1 C_1}}, \ \ \alpha_1 = \frac{1}{L_1 G_1}, \ \ \kappa_1 =  \frac{1}{\sqrt{M C_1}}, \ \  \gamma_1 = \frac{1}{M G_1}, \nonumber \\
\omega_2 = \frac{1}{\sqrt{L_2 C_2}}, \ \ \alpha_2 = \frac{1}{L_2 G_2}, \ \ \kappa_2 =  \frac{1}{\sqrt{M C_2}}, \ \  \gamma_2 = \frac{1}{M G_2}. \nonumber
\end{gather}

Eq. (\ref{eqn:IV_v_Laplace}) can be normalized by scaling the frequency $\tilde{\omega}=\omega/\omega_1$ as follows,
\begin{align}\label{eqn:IV_v_Laplace_normalized}
\left(- \tilde{\omega}^2 + \frac{j \tilde{\omega}}{ \tilde{\alpha}_1 } + 1\right) v_{1}(\tilde{\omega}) + \left(\frac{-\tilde{\omega}^2}{ \tilde{\kappa}_{2}^2} + \frac{j\tilde{\omega}}{\tilde{\gamma}_{2}} \right) v_{2}(\tilde{\omega}) =0, \\ 
\left(\frac{-\tilde{\omega}^2}{\tilde{\omega}_{2}^2} + \frac{j\tilde{\omega}}{\tilde{\alpha}_{2} } + 1\right) v_{2}(\tilde{\omega}) + \left(\frac{-\tilde{\omega}^2}{\tilde{\kappa}_{1}^2} + \frac{j\tilde{\omega}}{\tilde{\gamma}_{1}} \right) v_{1}(\tilde{\omega}) =0, \nonumber
\end{align}
with
\begin{gather}
\tilde{\alpha}_1 = \frac{1}{\tilde{G}_1}, \ \  \tilde{\kappa}_{1} = \frac{1}{\sqrt{\tilde{M}}}, \ \ \gamma_1 = \frac{1}{\tilde{M}  \tilde{G}_1}, \nonumber \\
\tilde{\omega}_2 = \frac{1}{\sqrt{\tilde{L}_2 \tilde{C}_2}}, \tilde{\alpha}_2 = \frac{1}{ \tilde{L}_2 \tilde{G}_2},  \tilde{\kappa}_{2} = \frac{1}{\sqrt{\tilde{M} \tilde{C}_2}}, \gamma_2 = \frac{1}{\tilde{M}  \tilde{G}_2}, \nonumber
\end{gather}
and the normalized quantities defined below,
\begin{gather}
 \tilde{G}_1 = \sqrt{\frac{L_1}{C_1}}G_1,  \ \ \tilde{M} = \frac{M}{L_1},  \nonumber \\
\tilde{L}_2 = \frac{L_2}{L_1}, \ \ \tilde{C}_2 = \frac{C_2}{C_1}, \ \ \tilde{G}_2 = \sqrt{\frac{L_1}{C_1}}G_2. \nonumber
\end{gather}

In terms of matrix form,Eq. (\ref{eqn:IV_v_Laplace_normalized}) can be expressed as follows,
\begin{equation}\label{eqn:IV_v_Laplace_matrix}
\overline{\overline{M}}(\tilde{\omega}) \begin{bmatrix}
v_{1}(\tilde{\omega})     \\
 v_{2}(\tilde{\omega}) \\
\end{bmatrix} =0, 
\end{equation} 
with
\begin{align}
\overline{\overline{M}}(\tilde{\omega})   =  \begin{bmatrix}
- \tilde{\omega}^2 + \frac{j \tilde{\omega}}{ \tilde{\alpha}_1 } + 1 & \frac{-\tilde{\omega}^2}{ \tilde{\kappa}_{2}^2} + \frac{j\tilde{\omega}}{\tilde{\gamma}_{2}}  \\
\frac{-\tilde{\omega}^2}{\tilde{\kappa}_{1}^2} + \frac{j\tilde{\omega}}{\tilde{\gamma}_{1}} &  \frac{-\tilde{\omega}^2}{\tilde{\omega}_{2}^2} + \frac{j\tilde{\omega}}{\tilde{\alpha}_{2} } + 1 \\   
\end{bmatrix} . \nonumber
\end{align}

\subsection{The Hamiltonian of the Coupled LRC Coil Resonators}
 From Eq. (\ref{eqn:IV_v_Laplace_matrix}), it can be seen that matrix equations contains second order frequency components. To cast the matrix equation into the Hamiltonian of a coupled resonators, two times of changes of basis are required.
 
 During the first basis change, $\overline{\overline{M}}(\tilde{\omega})$ is transformed such that the diagonal terms are reduced to zeros and the following transformation matrix $\overline{\overline{T}}_1$ is given below,
 \begin{align}
\overline{\overline{T}}_1   = \begin{bmatrix}
L_2 &   -M  \\
-M &  L_1 \\   
\end{bmatrix} .
\end{align}

Then a second change of basis is performed with the transformation matrix $\overline{\overline{T}}_2$ and the Hamiltonian equations of the coupled resonators can be expressed as follows,
 \begin{equation}\label{eqn:Hamiltonian}
\overline{\overline{T}}_1 \overline{\overline{M}}(\tilde{\omega})\begin{bmatrix}
v_{1}(\tilde{\omega})     \\
 v_{2}(\tilde{\omega}) \\
\end{bmatrix}  =  \left[ \tilde{\omega} \overline{\overline{I}} - \mathcal{H}(\tilde{\omega})    \right] \overline{\overline{T}}_2  \begin{bmatrix}
v_{1}(\tilde{\omega})     \\
 v_{2}(\tilde{\omega}) \\
\end{bmatrix} =0, 
\end{equation} 
where $\overline{\overline{I}}$ is the unit matrix and 
\begin{align}
 \mathcal{H}(\tilde{\omega})  = \begin{bmatrix}
 \tilde{\Omega}_1 &   \kappa_{1, 2}   \\
 \kappa_{2, 1} &  \tilde{\Omega}_{2}\\   
\end{bmatrix} ,
\end{align}
and the following transformation matrix,
\begin{align}\label{eqn:transform}
\overline{\overline{T}}_2   = \begin{bmatrix}
\tilde{\omega} - \tilde{\Omega}_1' &   \kappa_{1, 2}'   \\
 \kappa_{2, 1}' &  \tilde{\omega} - \tilde{\Omega}_{2}' \\   
\end{bmatrix} .  
\end{align}
 
Substituting Eq. (\ref{eqn:transform}) into Eq. (\ref{eqn:Hamiltonian}), the following is obtained,
 \begin{align}\label{eqn:Hamiltonian_transform}
\overline{\overline{T}}_1 \overline{\overline{M}}(\tilde{\omega})    = \left[ \tilde{\omega} \overline{\overline{I}} - \mathcal{H}(\tilde{\omega})    \right] \overline{\overline{T}}_2 ,
 \end{align}
 from which all the unknown parameters can be solved.
 
 \subsection{The Eigenvalues and Eigenvectors}
 With the obtained Hamiltonian $\mathcal{H}$, the eigenvalues of the coupled resonant coils are given by,
 \begin{align}\label{eqn:eigenvalues}
\tilde{\Omega}^\pm =  \frac{\tilde{\Omega}_1 + \tilde{\Omega}_2 }{2} \pm \sqrt{\kappa_{12} \kappa_{21} + \left(\frac{\tilde{\Omega}_1 - \tilde{\Omega}_2 }{2}\right)^2},
 \end{align}
 and the corresponding eigenvectors are as follows,
 \begin{align}\label{eqn:eigenvectors}
v^\pm=\left[\frac{\kappa_{12}}{ \tilde{\Omega}^\pm -  \tilde{\Omega}_1  }, 1 \right]^T.
 \end{align} 
 
 \subsection{Solutions of the Hamiltonian}
 The solution of the Hamiltonian $\mathcal{H}$ and the corresponding basis transform matrix $\overline{\overline{U}}$ can be obtained by solving Eq. (\ref{eqn:Hamiltonian_transform}), which are shown as follows,
 \begin{align}\label{eqn:Hamiltonian_solutions}
 \tilde{\Omega}_1^\pm = j \frac{G_1}{2} \pm  \tilde{\Omega} \left(\tilde{\Omega}_2^\pm \right),
 \end{align}
 with
 \begin{align}
 \left\{ \left[\tilde{\Omega} \left(\tilde{\Omega}_2^\pm \right) \right]^2 - \Gamma^2 \right\} \left\{ \left[\tilde{\Omega}^\pm \left(\tilde{\Omega}_2^\pm \right) \right]^2 - \Gamma'^2 \right\} + M^2 = 0, \nonumber 
 \end{align}
 and the following parameters,
 \begin{gather}
\tilde{\Omega} \left(\tilde{\Omega}_2^\pm \right) = \sqrt{\left( \tilde{\Omega}_2^\pm - j \frac{G_2}{2C_2} \right)^2 + (\Gamma^2- \Lambda^2)}, \nonumber \\
\tilde{\Omega}^\pm \left(\tilde{\Omega}_2^\pm \right) = \tilde{\Omega}_2^\pm \pm \tilde{\Omega} \left(\tilde{\Omega}_2^\pm \right) - j \frac{G_2}{2 C_2}, \nonumber \\
\Lambda = \frac{1}{2 C_2}  \sqrt{ \frac{4 C_2 - G_2^2 L_2 + G_2^ 2  M^2}{{L_2 - M^ 2}}} \nonumber \\
\Gamma =\frac{1}{2}  \sqrt{\frac{4 L_2-G_1^2L_2 + G_1^2 M^2 }{L_2 - M^2}}, \ \ \Gamma' = j \left(\frac{G_2}{2 C_2} - \frac{G_1}{2} \right).  \nonumber
 \end{gather} 
 Also, the coupling terms $\kappa_{12}$ and $\kappa_{21}$ are obtained as follows,
 \begin{align}
\kappa_{12} = \frac{M}{-j \frac{G_2}{C_2} + \tilde{\Omega}_1 + \tilde{\Omega}_2}; 
\kappa_{21} = \frac{\Gamma^2 - \left(\tilde{\Omega}_1 - j \frac{G_1}{2} \right)^2 }{\kappa_{12}}. 
 \end{align}
 
 Finally, other parameters such as $\tilde{\Omega}_1', \tilde{\Omega}_2', \kappa_{12}', \kappa_{21}'$ of the transformation matrix $\overline{\overline{T}}_2$ can be obtained accordingly.
 \begin{gather}
\tilde{\Omega}_1'^\pm = jG_1 - \tilde{\Omega}_1^\pm,  \tilde{\Omega}_2'^\pm =  \frac{C_2 M}{jG_2 -C_2 \tilde{\Omega}_1^\pm - C_2 \tilde{\Omega}_2^\pm}, \\
\kappa_{12}' = c_3 \tilde{\Omega}_1^{\pm3} + c_2 \tilde{\Omega}_1^{\pm2} + c_1 \tilde{\Omega}_1^{\pm} + c_0, \kappa_{21}' = \tilde{\Omega}_2^\pm - j\frac{G_2}{C_2}, \nonumber
 \end{gather} 
 with
 \begin{gather}
 c_1 = \frac{jG_1 \tilde{\Omega}_2^\pm}{M} + \frac{L_2}{M (L_2 - M^2)}+ \frac{G_1 G_2}{C_2M}, \nonumber \\
 c_2 = -\frac{\tilde{\Omega}_2^\pm}{M}+j\left( \frac{G_1}{M} + \frac{G_2}{C_2 M} \right), \nonumber \\
c_3 = - \frac{1}{M}, c_0 = L_2 \frac{C_2 \tilde{\Omega}_2^\pm - jG_2}{C_2 M (L_2 - M^2)}. \nonumber  
 \end{gather}

\section{Discussion}
Analytical solutions exists for some special cases. Also, the Hamiltonian can be normalized to give unit frequency of one coil resonator, \ie, $\omega_1 =1$.

\subsection{Identical Resonant Frequencies and Decay Rates}\label{subsec:equal_G1_G2}
When the two resonant coils have identical resonant frequencies and decay rates, the following are satisfied,
\begin{align}
\omega_1 = \frac{1}{\sqrt{L_1 C_1}} = \omega_2  = \frac{1}{\sqrt{L_2 C_2}};  \tau_1 = \frac{G_1}{C_1} = \tau_2 = \frac{G_2}{C_2}, \nonumber 
\end{align}
from which the Eq. (\ref{eqn:Hamiltonian_solutions}) has the following four solutions,
\begin{gather}\label{eqn:w1w2_G1G2}
 \tilde{\Omega}_1 =  \tilde{\Omega}_2 =  \pm \frac{\sqrt{\Gamma^2 \pm \sqrt{-M^2 + \Gamma^4}}}{\sqrt{2}} + j\frac{G_1}{2}. \nonumber
\end{gather}    

\subsection{Parity-Time Symmetry}\label{subsec:PT}
When the loss and gain of the resonant coils balance each other, the following are satisfied,
\begin{align}
\omega_1 = \frac{1}{\sqrt{L_1 C_1}} = \omega_2  = \frac{1}{\sqrt{L_2 C_2}};  \tau_1 = \frac{G_1}{C_1} = \tau_2 =- \frac{G_2}{C_2}, \nonumber 
\end{align}
from which the Eq. (\ref{eqn:Hamiltonian_solutions}) has the following six solutions,
\begin{gather}\label{eqn:w1w2_G1_G2_1}
 \tilde{\Omega}_1 =  - \tilde{\Omega}_2 = j \left(\frac{G_1}{2} \pm \frac{\sqrt{M^2 - G_1^2 \Gamma^2}}{G_1} \right), 
\end{gather}  
and
\begin{gather}\label{eqn:w1w2_G1_G2_2}
 \tilde{\Omega}_1 =   \tilde{\Omega}_2 + j G_1, \nonumber \\
 \tilde{\Omega}_2 =  
 \pm  \frac{\sqrt{\Gamma^2-\frac{G_1^2}{4} \pm \sqrt{-M^2 + \left(\Gamma^2 + \frac{G_1^2}{4}\right)^2}}}{\sqrt{2}}- j\frac{G_1}{2}. \nonumber  
\end{gather}   

\subsection{Identical Lossless Resonators Frequencies}
When the two resonant coils have identical resonant frequencies and lossless, 
\begin{align}
\omega_1 = \frac{1}{\sqrt{L_1 C_1}} = \omega_2  = \frac{1}{\sqrt{L_2 C_2}}; \ \ G_1 = G_2 =0, \nonumber 
\end{align}
and the following solutions are obtained,

\begin{align}
 \tilde{\Omega}_1 =  \tilde{\Omega}_2 =  \pm\sqrt{\frac{L_2 \pm \sqrt{ L_2^2  -M^2(M^2 - L_2)^2}}{2 (L_2 - M^2)}},
\end{align}
and
\begin{gather}
\kappa_{12} = \kappa_{21} = \frac{M}{2 \tilde{\Omega}_1}, \nonumber
\end{gather}
which gives the the eigenvalues and eigenvectors according to Eq. (\ref{eqn:eigenvalues}) and Eq. (\ref{eqn:eigenvectors}) as follows,
\begin{align}
\tilde{\Omega}^\pm = \tilde{\Omega}_1 \pm \kappa_{12}, \ \ v^\pm=\left[\pm 1, 1 \right]^T,
 \end{align}
agreeing with symmetric/anti-symmetric eigenmodes due to the off-diagonal coupling.


\begin{figure}[ht]
\centering
  \includegraphics[width=1\linewidth]{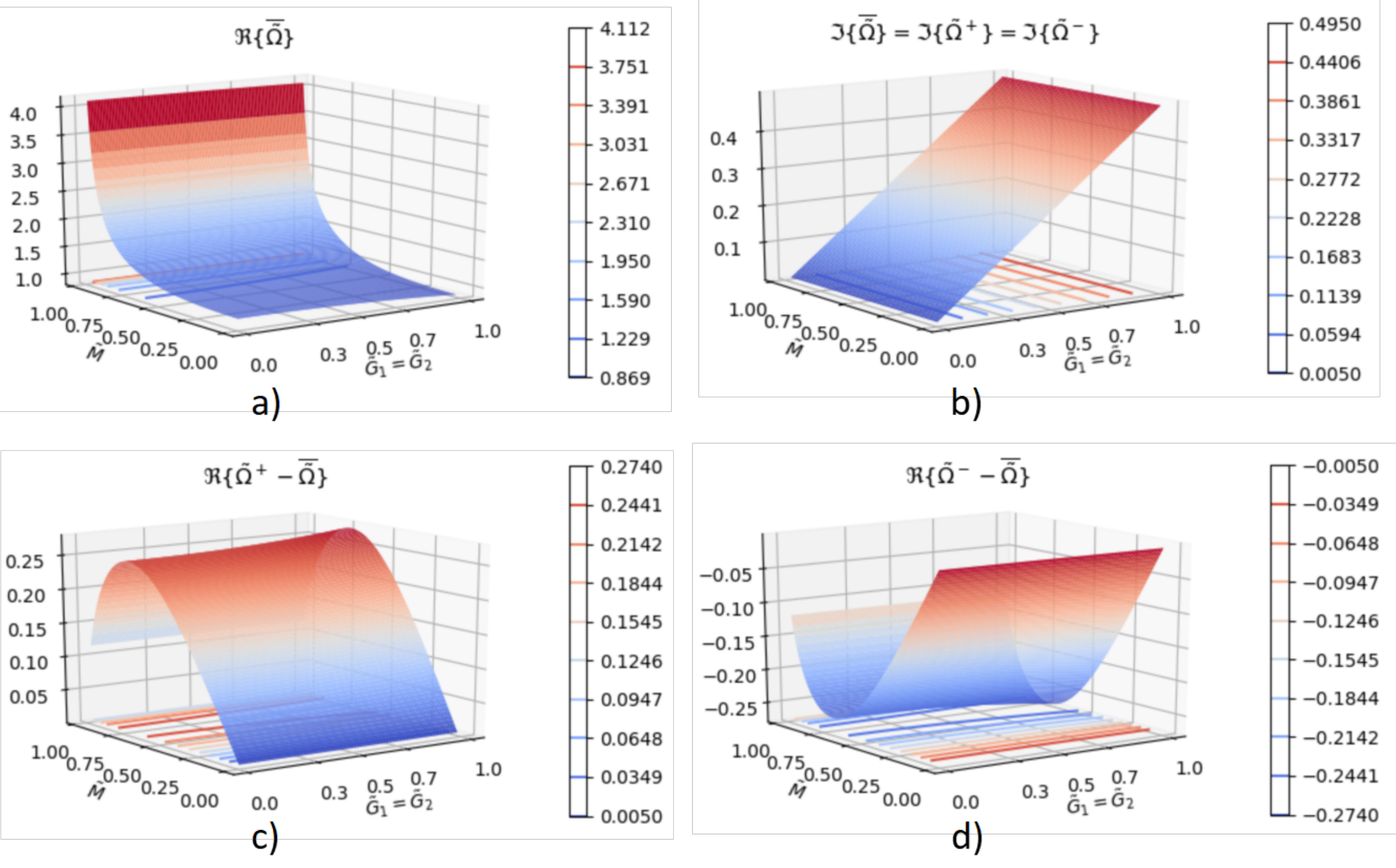} 
  \caption{Simulation result for the coupled coil resonators pair with identical resonant frequencies and decay rates: a) the real part of the mean of the two eigenfrequencies; b) the imaginary part of the mean of the two eigenfrequencies; c) the deviation of the real part of eigenfrequency \#1 from the real part of the eigenfrequency mean; and d) the deviation of  the real part of eigenfrequency \#2 from the real part of the eigenfrequency mean.}
  \label{fig:equal_G1_G2}
\end{figure}  

\begin{figure}[ht]
\centering
  \includegraphics[width=1\linewidth]{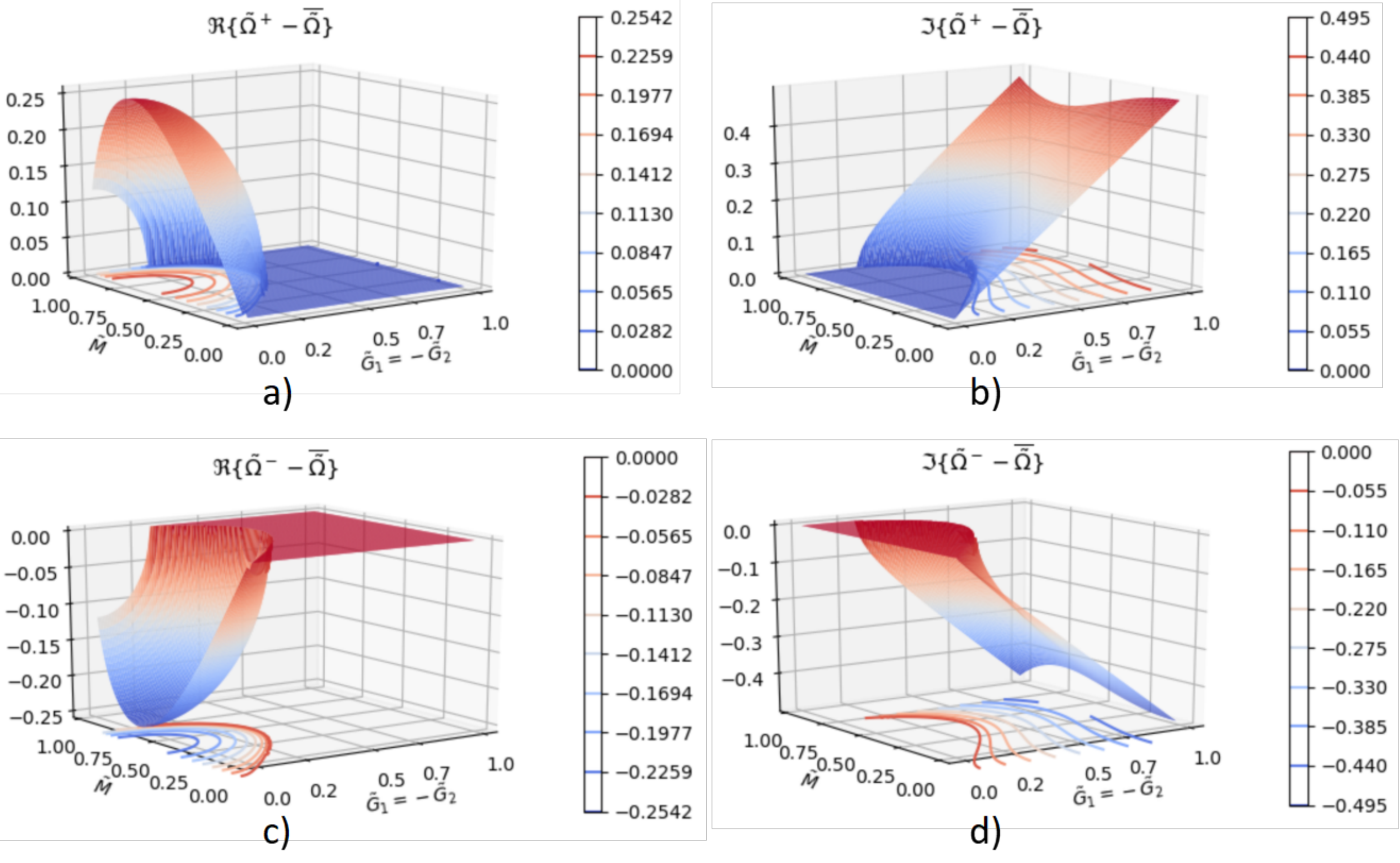} 
  \caption{Simulation result for the PT-symmetric coupled resonator pair with balanced gain and loss: a) the deviation of the real part of eigenfrequency \#1 from the eigenfrequency mean; b) the deviation of the imaginary part of eigenfrequency \#1 from the eigenfrequency mean; c) the deviation of the real part of eigenfrequency \#2 from the eigenfrequency mean; d) the deviation of the imaginary part of eigenfrequency \#2 from the eigenfrequency mean.}
  \label{fig:PT}
\end{figure} 

\section{Simulation Results}
Numerical simulation has been performed to confirm the correctness of the quantum formulation. Simulation  results of two typical cases are shown here: 1) identical resonant frequencies and decay rates case as shown in Section \ref{subsec:equal_G1_G2}; and 2) PT symmetry case as shown in Section  \ref{subsec:PT}. 

Fig. \ref{fig:equal_G1_G2} shows the surface and contours plots for case 1) with $\omega_1 = 1/\sqrt{L_1 C_1} = \omega_2 =  1/\sqrt{L_2 C_2} =1$ and $G_1 = G_2$: a) real part of the mean of the two eigenfrequencies, \ie, $\Re\{\overline{\tilde{\Omega}} \}= ( \Re\{\tilde{\Omega^+} \} + \Re\{ \tilde{\Omega^-})/2 \}$; b) imaginary part of the mean of the two eigenfrequencies, \ie, $\Im\{\overline{\tilde{\Omega}} \}= ( \Im\{\tilde{\Omega^+} \} + \Im\{ \tilde{\Omega^-})/2 \}$; c) deviation of real part of one eigenfrequency from the real part of the two eigenfrequencies' mean, \ie, $\Re\{\tilde{\Omega}^+\} - \Re\{\overline{\tilde{\Omega}} \}$; and c) deviation of real part of the other eigenfrequency from the real part of the two eigenfrequencies' mean, \ie, $\Re\{\tilde{\Omega}^-\} - \Re\{\overline{\tilde{\Omega}} \}$.  Fig. \ref{fig:equal_G1_G2}a) shows that the two eigenfrequencies' mean value increases for an increasing coupling coefficient $M$. Also, Fig. \ref{fig:equal_G1_G2}b) shows an increasing imaginary part of the mean eigenfrequency when the losses of the coil resonators $G_1 = G_2$ increase. Note that the imaginary parts of the two eigenfrequencies are identical due to symmetry. Finally, Fig. \ref{fig:equal_G1_G2}c) Fig. \ref{fig:equal_G1_G2}d) show the deviation of the real parts of the two eigenfrequencies from that of the eigenfrequency mean: it is clear that the deviations are anti-symmetric due to symmetry and repelling coupling effect.

Fig. \ref{fig:PT} shows the surface and contour plots for PT-symmetry case 2) with $\omega_1 = 1/\sqrt{L_1 C_1} = \omega_2 =  1/\sqrt{L_2 C_2} =1$ and $G_1 = -G_2$: a) deviation of the real part of the  eigenfrequency \#1 from that of the eigenfrequency mean, \ie, $\Re\{\tilde{\Omega}^+\} - \Re\{\overline{\tilde{\Omega}} \}$; and b) deviation of the imaginary part of the  eigenfrequency \#1 from that of the eigenfrequency mean, \ie, $\Im\{\tilde{\Omega}^+\} - \Im\{\overline{\tilde{\Omega}} \}$; c) deviation of the real part of the  eigenfrequency \#2 from that of the eigenfrequency mean, \ie, $\Re\{\tilde{\Omega}^-\} - \Re\{\overline{\tilde{\Omega}} \}$; and d) deviation of the imaginary part of the  eigenfrequency \#2 from that of the eigenfrequency mean, \ie, $\Im\{\tilde{\Omega}^-\} - \Im\{\overline{\tilde{\Omega}} \}$. Comparing Fig. \ref{fig:PT}a) to Fig. \ref{fig:PT}c), and  Fig. \ref{fig:PT}b) to Fig. \ref{fig:PT}d), it is clear that the deviation of both the real part and imaginary part of the eigenfrequencies are anti-symmetric, due to the anti-symmetry of the PT-symmetric gain/loss profile. What's more important, Fig. \ref{fig:PT}a) to Fig. \ref{fig:PT}d) clearly show the evolution of the real eigenfrequencies to the imaginary eigenfrequencies when the gain/loss $G_1 = - G_2$ becomes larger than the coupling coefficient $M$, confirming the Exception Points (EPs) of the PT-symmetric system. At last, when the gain/loss is large enough, no real eigenfrequencies exist for such non-Hermitian Hamiltonian system.

\section{Conclusion}\label{sec:con}
Quantum formulation of the PT symmetric coupled LRC coil resonators system has been derived. Starting from the governing equation system, two basis transforms are performed to transform the second order frequencies equations system in the frequency domain to a quantum Hamiltonian and the corresponding new basis. The first basis transform is to make off-diagonal terms of  the equations system to be constant values that contain no frequency term. Then a second basis transform is performed to obtain the quantum Hamiltonian with the off-diagonal effective complex frequencies and the off-diagonal coupling terms, together with the new basis that denotes the coupled quantum states. With the obtained quantum Hamiltonian, eigenvalues and the corresponding eigenvectors can be obtained as usual. Finally, numerical simulation confirms the correctness of the theory. It is expected that the quantum formulation of the coupled coil resonators provide helpful insight and guideline for the PT symmetric RF/microwave resonators and systems.


\end{document}